\documentclass[prl,twocolumn,amssymb,amsmath,superscriptaddress]{revtex4}
\usepackage{hyperref}
\usepackage{graphicx}
\usepackage{color}

\usepackage{amsmath}

\def\one{{\mathchoice {\rm 1\mskip-4mu l} {\rm 1\mskip-4mu l} {\rm \mskip-4.5mu l} {\rm 1\mskip-5mu l}}}

\def\NOT(#1,#2){\OneQubitGate(#1,#2){$X$}}

\begin{document}

\title{Digital Quantum Simulation of the Statistical Mechanics of a Frustrated Magnet}


\author{Jingfu Zhang}
\thanks{These authors contributed equally to this work.}
\affiliation{Institute for Quantum Computing, University of
Waterloo, Waterloo, Ontario, N2L 3G1, Canada}
\affiliation{Department of Physics and Astronomy, University of
Waterloo, Waterloo, Ontario, N2L 3G1, Canada}

\author{Man-Hong Yung}
\thanks{These authors contributed equally to this work.}
\affiliation{Department of Chemistry and Chemical Biology, Harvard
University, Cambridge, MA, 02138, USA}

\author{Raymond Laflamme}
\affiliation{Institute for Quantum Computing, University of
Waterloo, Waterloo, Ontario, N2L 3G1, Canada}
\affiliation{Department of Physics and Astronomy, University of
Waterloo, Waterloo, Ontario, N2L 3G1, Canada}
\affiliation{Perimeter Institute for Theoretical Physics,
Waterloo, Ontario, N2J 2W9, Canada}

\author{Al\'{a}n Aspuru-Guzik}
\affiliation{Department of Chemistry and Chemical Biology, Harvard University, Cambridge, MA, 02138, USA}

\author{Jonathan Baugh}
\affiliation{Institute for Quantum Computing, University of Waterloo, Waterloo, Ontario, N2L 3G1, Canada}
\affiliation{Department of Chemistry, University of Waterloo, Waterloo, Ontario, N2L 3G1, Canada}
\affiliation{Department of Physics and Astronomy, University of Waterloo, Waterloo, Ontario, N2L 3G1, Canada}

\date{\today}

\pacs{03.67.Lx}

\begin{abstract}
Many interesting problems in physics, chemistry, and computer science are equivalent to problems of interacting spins \cite{Mezard1987}. However, most of these problems require computational resources that are out of reach by classical computers. A promising solution to overcome this challenge is to exploit the laws of quantum mechanics to perform simulation \cite{Feynman_1982}. Several ``analog" quantum simulations of interacting spin systems have been realized experimentally \cite{Peng2005,Friedenauer2008,Kim2010,Edwards2010,Ma2010,Struck2011}. However, relying on adiabatic techniques, these simulations are limited to preparing ground states only. Here we report the first experimental results on a ``digital" quantum simulation on thermal states; we simulated a three-spin frustrated magnet, a building block of spin ice, with an NMR quantum information processor, and we are able to explore the phase diagram of the system at any simulated temperature and external field. These results serve as a guide for identifying the challenges for performing quantum simulation on physical systems at finite temperatures, and pave the way towards large scale experimental simulations of open quantum systems in condensed matter physics and chemistry.
\end{abstract}



 \maketitle
The most challenging aspect of many-body simulation is that the memory and temporal resources often scale exponentially, rendering many problems of interest intractable by all known classical methods  \cite{Feynman_1982}. A promising solution is \textit{quantum simulation}, in which a quantum system acts as a processor to simulate another physical system (quantum or classical). There are two classes of quantum simulation: ``analog" simulators are typically engineered to simulate a particular class of Hamiltonians~\cite{Buluta_Nori_2009} and to find ground states of non-trivial Hamiltonians adiabatically, whereas ``digital" simulators rely on universal quantum information processors (QIPs), capable of implementing a universal set of quantum gate operations \cite{Kassal2010} to simulate not only ground states, but also thermal states, and even time-evolving states.

Simulations of interacting spin systems are of particular importance to many applications, such as modeling magnetism \cite{Mattis2008}, solving optimization problems \cite{Young2008}, and restoring digital image \cite{Nishimori1999}. Furthermore, understanding the properties of the spin models also offers insights to the computational complexity theory~\cite{Istrail2000}. For example, the ground-state problem of the Ising spin model is known to be an $\mathsf{NP}$-complete problem; this implies that if an efficient algorithm for solving the ground-state problem of the Ising model exists, then it can solve all other problems in the class of $\mathsf{NP}$. This matter is related to the question whether $\mathsf{P}$ equals $\mathsf{NP}$, and is a major unsolved problem in computer science.

In a series of recent experiments \cite{Peng2005, Friedenauer2008, Kim2010, Edwards2010}, based on the adiabatic methods, progresses of quantum simulation of various spin systems have been achieved in various physical systems. These experiments, however, suffers from two limitations: (a)~they are limited to studying the ground-state properties only, and (b)~the energy gaps along the adiabatic paths must be large enough to avoid excitations from the ground states. In general, the energy gaps cannot be pre-determined efficiently, and are non-controllable. Therefore, the advantage of the adiabatic methods over classical methods is not guaranteed for all cases.

On the other hand, at finite temperatures, all of the thermodynamics of spin systems can be obtained by determining the partition function $\mathcal{Z}$, which, instead of $\mathsf{NP}$, falls into a different complexity class called sharp-$\mathsf{P}$, or $\mathsf{\#P}$. However, if an efficient algorithm for evaluating partition functions exists, then the ground-state properties of the corresponding spin systems can also be determined efficiently. Therefore, the problem of determining partition functions is at least as hard as the $\mathsf{NP}$-problems, or simply called $\mathsf{NP}$-hard.

Practically, partition functions cannot be computed efficiently, except for some simple cases such as 1-D spin chains. For classical spins, the classical Metropolis algorithm provide a means for generating the Gibbs distributions, through the construction of Markov chains with Monte-Carlo methods. For quantum systems, the quantum generalization of the Metropolis algorithm has been achieved \cite{Temme2011, Yung2011}. However, Markov-chain based methods, similar to the adiabatic methods, are limited to the cases where the Markov-matrix gaps cannot be too small to achieve convergence. Particularly, for frustrated spin systems, Metropolis sampling can result in ensembles trapped in local minima. In these cases, methods for direct encoding the Gibbs distribution into the states of the qubits would be more efficient. This is the key issue that motivates this experimental work.

By using a digital simulator, our system can explore the full phase diagram of the thermal state of the system, as a function of temperature and magnetic field. All ranges are in principle accessible due to the digital nature of the simulation. Unlike other quantum phase transitions explored in other quantum simulation experiments \cite{Friedenauer2008,Kim2010,Edwards2010}, the underlying Hamiltonian of the simulator does not need to be restricted to certain types of interactions to be able to simulate the target system.

\begin{figure}[t]
\includegraphics[width=0.9 \columnwidth]{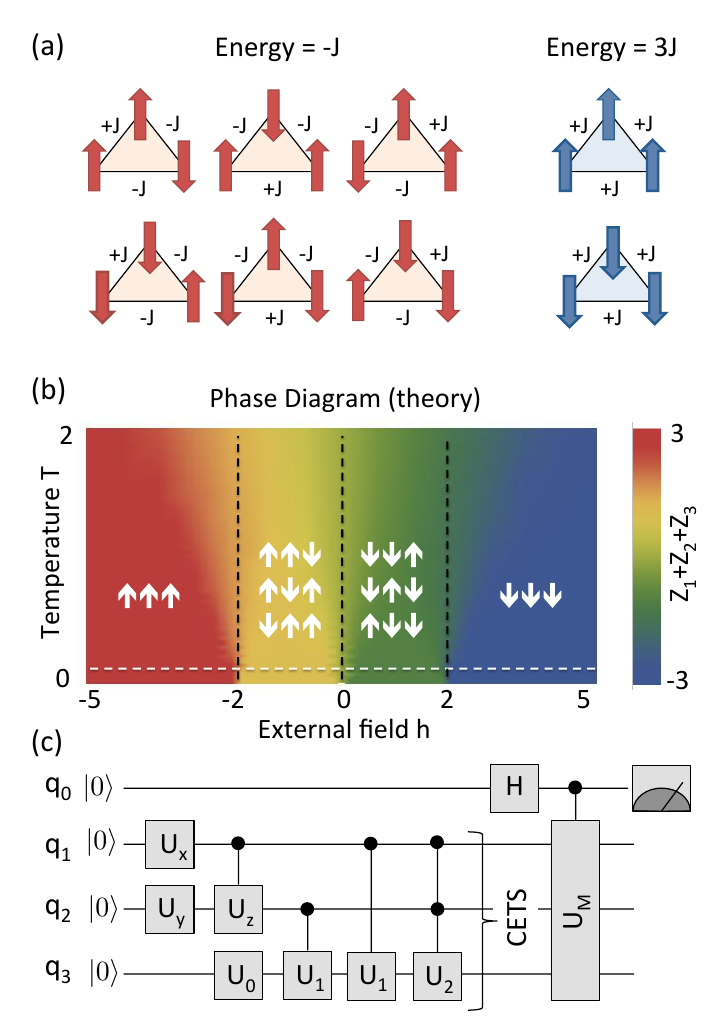} 
\caption{Theoretical descriptions of the frustrated magnet. (a)
All possible configurations of a three-spin frustrated magnet at
zero temperature and zero magnetic field. There is a six-fold
degeneracy in the ground state, leading to a non-zero entropy.
 (b) Theoretical phase diagram.  The units of the axes are $k_{B}$ and $J$ for temperature and external field.
 The dashed line parallel $h$- axis denotes $T = 1/11$, which corresponds to the case of the experimental
data taken in fig. 3. (c) Quantum circuit diagram for preparing
and measuring the CETS $|\Psi_{\beta}\rangle$ defined in Eq.
\ref{CETS} from the initial state $\left| {0000} \right\rangle$.
 The gate sequence for preparing the CETS on the lower
three qubits is determined by the method in Ref. \cite{Yung2010}.
An explicit way for constructing the $U_k$-gates is given in the
supplementary materials. The top qubit $q_0$ serves as a probe for
measuring the physical observables. } \label{figcir}
\end{figure}

In this letter, we report the first digital quantum simulation of the finite-temperature properties
a classical three-spin frustrated magnet, a building block of spin ice, (see Fig.~\ref{figcir}),
 using a four-qubit quantum register based on NMR. The reason for simulating the
 frustrated magnet is that it exhibits a rich phase diagram of the total magnetization as a
 function of temperature and magnetic field. This allows us to experimentally probe various
 distinct features of this system. On the other hand, the phenomenon
 of geometric frustration is an interesting topic in condensed matter
 physics. For example, materials, such as water ice, exhibiting geometric
 frustration cannot be completely frozen; the motion at the molecular scale continues even at absolute zero.
 Recently, the same three-spin frustrated magnet at zero temperature has been
 simulated by trapped ions \cite{Kim2010,Edwards2010}.  We aim to make progress
 along this direction by extending the quantum simulation of the frustrated magnet to finite temperatures.

In our simulation, instead of a mixed state, the implemented algorithm prepares a coherent encoding of a classical thermal state (CETS) on a quantum register~\cite{Lidar1997, Yung2010},
\begin{equation}\label{CETS}
    |\Psi_{\beta}\rangle = \sum_{k}\sqrt{e^{-\beta E_{k}}/\mathcal{Z}}|\phi_k\rangle \quad ,
\end{equation}
which is a pure state (a pseudopure state in the NMR experiment)
with amplitudes $\sqrt {e^{ - \beta E_k } /\mathcal{Z}}$ equal to
the square roots of the corresponding thermal state Gibbs
distribution associated with the eigenstate $\left| {\phi _k }
\right\rangle$ of the Hamiltonian $H$. Here $\beta =1/T$
($k_B=1$), and $\mathcal{Z} = {\rm Tr} ( {e^{ - \beta H} })$ is the
partition function. The CETS, therefore, contains all of the
information about the thermal density matrix
\begin{equation}\label{thermal}
    \rho_{th} = e^{-\beta H}/\mathcal{Z}
\end{equation}
of the system. In fact, the thermal density matrix $\rho_{th}$ can be directly obtained from the CETS state
 $|\Psi_{\beta}\rangle$ by artificially ``decohering" the off-diagonal elements of the density matrix
  $|\Psi_{\beta}\rangle\langle \Psi_{\beta}|$ constructed from the CETS.

In this method \cite{Yung2010}, the number of quantum gates needed
to prepare the CETS is linear in the number of spins for 1D cases
and
sub-exponential for 2D cases, but is still exponential in
general for $\mathsf{NP}$-problems. Nonetheless, the efficiency
of this algorithm is independent of the simulated temperature, and not limited by the small-gap problem encountered in the Markov-Chain Monte Carlo algorithms. This makes it advantageous for simulating the low-temperature properties of frustrated spin systems. Furthermore, although this algorithms can at most yield a quadratic speedup for simulating the most general thermal states \cite{Yung2010}, the subclass of the CETS which can be created efficiently on a quantum computer can serve as a ``heat bath" \cite{Winograd2009} for the simulation of the dynamics of open quantum systems, which could give an exponential advantage \cite{Lloyd1996}. Our goal is to investigate how well such a CETS can be prepared in the laboratory subject to the existing experimental constraints. 

In this experiment, three qubits encode the CETS of a triangle
plaquette of Ising spins with equal couplings~$J$, with
temperature $T$ and global magnetic field $h$ as variables. A
fourth ancilla qubit is used to probe the physical properties of
the CETS by measuring the set of diagonal Pauli operators, so that
quantities such as the total magnetizations and spin correlations can
be extracted. These measurements are sufficient for determining
the partition function $\mathcal{Z}$, from which any thermodynamic
quantity of interest, e.g. entropy $S$, can be calculated.

To be more specific, the Hamiltonian of the frustrated magnet is defined by
\begin{equation}\label{magnet}
H = J\left( {Z_1 Z_2  + Z_2 Z_3 + Z_1 Z_3 } \right) + h\left( {Z_1
+ Z_2  + Z_3 } \right) \, ,
\end{equation}
where $Z_i \equiv \sigma^{z}_{i}$ is the $\hat{z}$ Pauli matrix of
the spin $i$. For $J{>}0$, the coupling is
antiferromagnetic, where the spins tend to minimize the energy by pointing in opposite directions. The external field $h$, however, tends to force the spins to align. 
The effect of finite temperature $T$ is to wash out both tendencies. 
The competitions between these factors give rise to a phase diagram with a rich structure, as shown in Fig. \ref{figcir}b. For example, near some critical values of the external fields $h =-2 J$, $0$ and $2 J$, there are crossover points where the configuration of the spins, and hence the total magnetization, change abruptly. Near $T=0$  and $h=0$, the ground state is fully frustrated with a six-fold degeneracy,  as illustrated in Fig.~\ref{figcir}a. This means that, unlike ordinary materials, the entropy (and hence heat capacity) of the frustrated spin system is non-zero at $T=0$.

Our strategy for the study is as follows: for any given value of
the temperature $T$ and magnetic field $h$, the CETS
$|\Psi_{\beta}\rangle$ can be prepared with a quantum circuit of
constant depth, as shown in Fig. \ref{figcir}c. Here we choose $J$
as the unit for $h$.  The three lower qubits, $q_1, q_2$ and
$q_3$, initialized into the $|000\rangle$ state, are chosen as
register qubits to encode the CETS. We choose the phase kick-back
method~\cite{Kaye2007} to extract information about the CETS by
introducing a fourth qubit $q_0$ as a probe qubit, shown as the
top qubit in Fig. \ref{figcir}c. The probe qubit is then prepared
in a superposition state $(|0\rangle+|1\rangle)/\sqrt{2}$ and a
controlled-$U_M$ gate operation is applied to the joint probe-CETS
system to measure observables $\langle U_M\rangle = \langle
\Psi_\beta |U_M|\Psi_\beta \rangle$ on the top ``probe" qubit.
Here $\langle U_M\rangle$ is proportional to the coherent element
in the reduced density matrix of the probe qubit,
\begin{equation}\label{sreduced}
\rho_{0}=\frac{1}{2}\left( \begin{array}{cc}
  1 & \langle U_M\rangle \\
  \langle U_M\rangle^{*} & 1 \\
\end{array}\right ) \quad ,
\end{equation}
through tracing over the register qubits and can be extracted from
the NMR signal of the probe qubit. By measuring the set of
operators
\begin{equation}\label{setU}
U_M  = \left\{ {Z_1 ,Z_2 ,Z_3 ,Z_1 Z_2 ,Z_2 Z_3 ,Z_1 Z_3 ,Z_1 Z_2
Z_3 } \right\} \, ,
\end{equation}
the full thermal
state density matrix $\rho_{th}$ can be reconstructed from the CETS.

\begin{figure}[t]
\includegraphics[width= 0.9 \columnwidth]{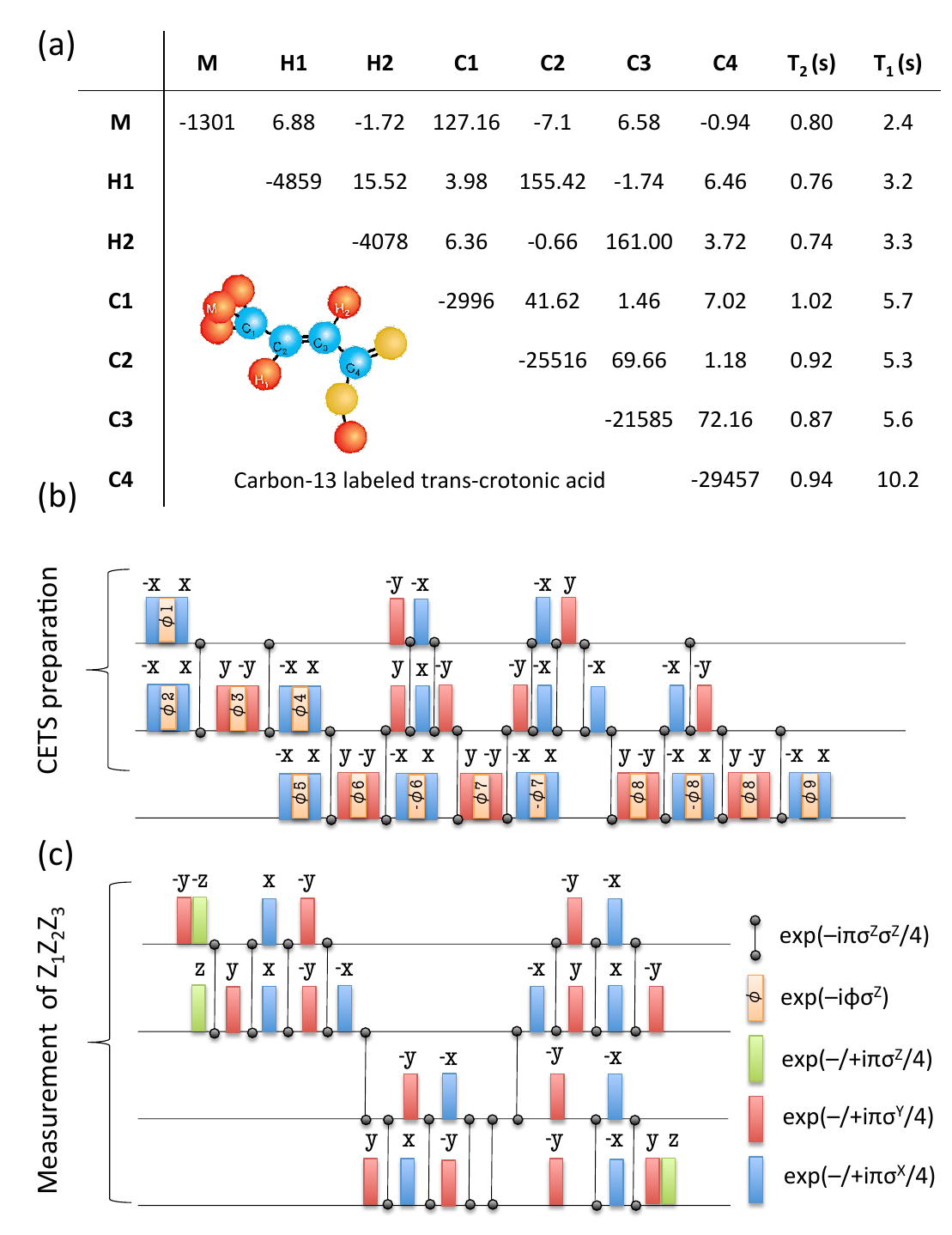} 
\caption{Experimental protocol. (a) Hamiltonian parameters for the
nuclear spins in carbon-13 labelled trans-crotonic acid, with the
structure shown as the inset. The chemical shifts and J-coupling
constants (in Hz) are listed on and above the diagonal in the
table, respectively. The longitudinal and transversal relaxation
times $T_1$ and $T_2$ measured by standard inversion recovery and
Hahn echo pulse sequences are listed at right. The chemical shifts
are given with respect to reference frequencies of 700.13 MHz
(protons) and 176.05 MHz (carbons). The molecule provides seven
qubits since the methyl group can be treated as a single qubit
using a gradient-based subspace
selection~\cite{Knill:2000fk}. 
(b,c) Pulse sequences for preparing the CETS and measuring $Z_1
Z_2 Z_3$ via the probe qubit, respectively, where refocusing
pulses are not shown. The three carbons
 C$_{2}$, C$_{3}$ and C$_{4}$ act as the CETS register qubits $1$, $2$ and
 $3$, and C$_{1}$ acts as the probe qubit. The rotation angles are determined by the angles in Fig. \ref{figcir} are listed in the supplementary
 material.} \label{figpulse}
\end{figure}

For the NMR implementation, we choose as a sample of the $^{13}$C-
labelled trans-crotonic acid dissolved in d6-acetone, which forms
a seven-qubit register; the four qubits in this experiment
corresponds to the four carbon spins, and the other three nuclear
spins are not directly involved after the preparation of the
pseudopure state. The experiments were carried out on a Bruker DRX
700 MHz spectrometer. The structure of the molecule and the
Hamiltonian parameters of the seven spin qubits are shown in Fig.
\ref{figpulse}a, where the NMR Hamiltonian of this system is given
by
\begin{equation}\label{ham}
    H_{NMR}=-\pi\sum_{i} \nu_{i}\sigma^{z}_{i}
  + \pi \sum_{k<l} J_{kl}\sigma^{z}_{k}\sigma^{z}_{l}/2 \quad ,
\end{equation}
where $\nu_{i}$ denotes the chemical shift of spin $i$, and
$J_{kl}$ denotes the coupling strength between spins $k$ and $l$.

In the experiment, we exploit standard Isech and Hermite-shaped
pulses to implement single-spin operations for the nuclei M and C$_1$-C$_4$, and
numerically optimized GRAPE pulses \cite{Khaneja2005296,Ryan2008}
for manipulating H$_1$ and H$_2$ (for initial pseudopure state preparation only). 
A custom-built software compiler generates pulse sequences, including refocussing pulses,
that are optimized for the highest unitary fidelity \cite{Ryan2008}.
Furthermore, the radio-frequency (r.f.) spin selection techniques are exploited to
improve the linewidth, and therefore the coherence, of the
ensemble qubits \cite{Maffei1991382,Knill:2000fk}. The effect of
pulse errors due to r.f. inhomogeneity is reduced by a spatial
selection of molecules in a region of high r.f.~homogeneity. A
labelled pseudo-pure state is prepared of the form
$\rho_{s}=\mathbf{0}\mathbf{0}\sigma_z
\mathbf{0}\mathbf{0}\mathbf{0}\mathbf{0}$  using methods described
in Ref.  \cite{Knill:2000fk}, where
$\mathbf{0}\equiv|0\rangle\langle0|$ and the order of qubits is as follows: M,
H$_1$, H$_2$,  C$_1$, C$_2$, C$_3$, C$_4$. Note that we employ the
deviation density matrix formalism \cite{Chuang98}. 

The four carbon spins, initialized in the state $\mathbf{0000}$,
are used to prepare and measure the CETS, where C$_1$ is the probe
qubit, and C$_2$ - C$_4$ are the register qubits for simulating
the frustrated magnet. The CETS $|\Psi_{\beta}\rangle$ is prepared
by the pulse sequence shown in Fig. \ref{figpulse}b. The NMR
signal of C$_1$ is acquired after the controlled-$U_M$ gate is
applied. The controlled-$U_M$ is implemented by combining
phase-flip and SWAP gate operations, and can be further be
decomposed into nearest-neighbor coupling evolutions and single
spin rotations. Fig. \ref{figpulse}c illustrates the sequence for
the observable $Z_1 Z_2 Z_3$.

\begin{figure}[t]
\includegraphics[width= 1 \columnwidth]{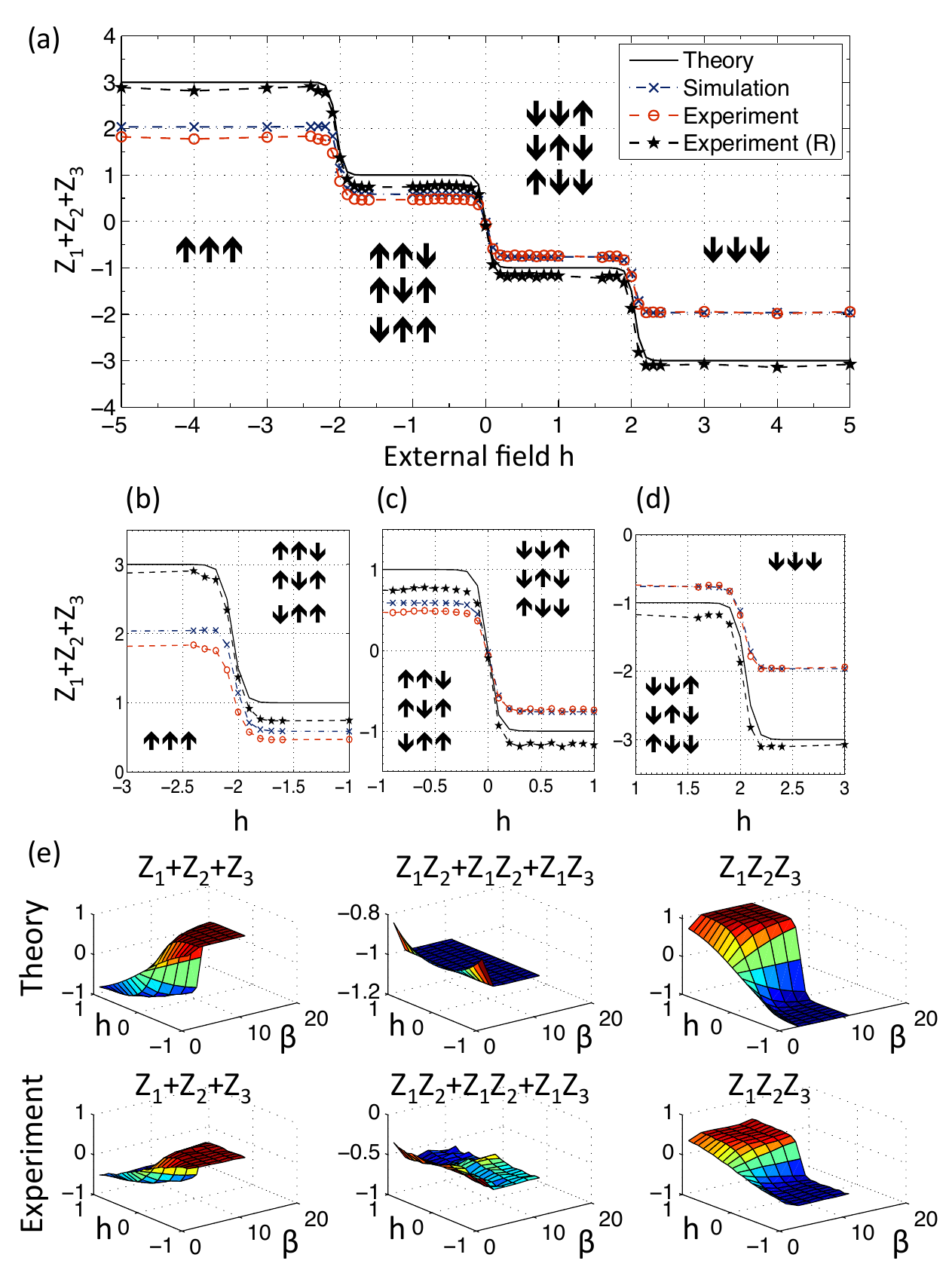} 
\caption{Experimental measured magnetization and correlations. (a)
Magnetization $Z_1+Z_2+Z_3$ as a function of magnetic field $h$ at
low temperature, $T = 1/11$. The experimental data (``$\circ$") is
plotted together with numerical simulation results (``$\times$")
that include effects of carbon $T_2$ and proton $T_1$. The
theoretical result is shown as the solid curve. The points labeled
``Experiment (R)" are obtained from the experimental data by using
a simple decoherence model to partially remove the effects of
decoherence with no free parameters (described in the text). The
sharp change show the phase transitions, and the regions around
the critical points $h=-2$, $0$, $2$ are enlarged as figures
(b-d), respectively.(e) Surface plots for the total magnetization
$Z_1+Z_2+Z_3$, correlations $Z_1 Z_2 + Z_2 Z_3 + Z_1 Z_3$, and
$Z_1 Z_2 Z_3$ in theory (top row) and measured in experiment
(bottom row). } \label{figmag}
\end{figure}

As indicated in Eq. (\ref{sreduced}), $\langle U_M\rangle$ is
encoded in the coherent part of the probe qubit (C$_1$) state. In
the spectra of probe qubit (see supplementary
 material), the coherence is distributed among
$2^6=64$ peaks, each of which corresponds to a particular
eigenstate of the remaining $6$ qubits M, H$_1$, H$_2$,
C$_2$-C$_4$. The intensities of these peaks are obtained by a
precise spectral fitting procedure \cite{Alex10}. In preparing and
measuring the CETS, no computational operations are performed on
the proton spins M, H$_1$, and H$_2$. The numerical simulations of
the experiment take into account the effect of $T_1$ relaxation
process of the proton spins, which is experimentally measured
though the decay of their initial state
$\mathbf{0}\mathbf{0}\sigma^{z}\one\one\one\one$, where $\one$
denotes the identity operator (see supplementary material for
details).

Since $\sigma^{z}=\mathbf{0}-\mathbf{1}$ with
${\mathbf{1}} \equiv |1\rangle\langle1|$, the $64$ peaks are divided into
two antiphase multiplets corresponding to the two eigenstates of
 H$_2$ (see supplementary materials). We may, for example, choose the group marked by the H$_2$ state $\mathbf{0}$.
By adding the intensities of the $8$
 peaks marked by the state $\mathbf{0}_M\mathbf{0}_{H_{1}}\mathbf{0}_{H_{2}}$,
 $\langle U_M\rangle$ is obtained, taking into account
 proper normalization relative to the initial pseudopure
 state $\rho_{s}$.

The experimental results are summarized as follows: the diagonal
elements of the density matrix constructed by the CETS are
determined by measuring the full set of diagonal Pauli operators
[see Eq.~(\ref{setU})] for a range of simulated temperatures $T$
and external fields $h$. The experimental results are shown in
Fig. \ref{figmag}. In Fig.\ref{figmag}a, for a low temperature
($\beta=11$), the total magnetization $Z_1+Z_2+Z_3$ of the
frustrated magnet is probed for a range of the simulated field
$h$. The raw experimental results are in good agreement with the
numerical simulation which takes into the account the decoherence
effects. By rescaling the total magnetization by a constant
factor, which is equivalent to removing the isotropic errors
\cite{Yung2011a} (see also supplementary material), the rescaled
results agree much better with the theoretical predictions. In any
case, it is clear that the magnetization changes in steps when the
simulated magnetic field is varied from a large negative value
($h=-5$) to a large positive value ($h=5$). The critical points
for the crossovers are located at $h=\pm2$, and $0$, in agreement
with the theoretically prediction shown in Fig. \ref{figcir}a.
This is the result of the competition between the
antiferromagnetic couplings and the external field applied to the
frustrated magnet.

Apart from the total magnetization, the other correlation functions are probed systematically for a range of the simulated temperatures and external fields. The results are shown in Fig.~\ref{figmag}e. From these data, we can construct the thermal state density matrix of the frustrated magnet (subject to the normalization condition ${\rm Tr}(\rho _{th}) = 1$):
\begin{equation}
\rho _{th}  = \frac{1}{8}\one + \sum\limits_i {a_i Z_i }  +
\sum\limits_{j < k} {b_{jk} } Z_j Z_k  + cZ_1 Z_2 Z_3 \quad,
\end{equation}
where $a_i  {\equiv} \left\langle {Z_i } \right\rangle /8$, $b_{j
k}  {\equiv} \left\langle {Z_j Z_k} \right\rangle /8$, and $c
{\equiv} \left\langle {Z_1 Z_2 Z_3} \right\rangle /8$. We ignore the imaginary parts of the elements, which are zero in theory, and less than 11\% in the experimental data. With complete knowledge of the thermal density matrix $\rho_{th}$,
we can determine  all of the macroscopic thermodynamic observables
for an ensemble of frustrated magnets. In this study, we are
particularly interested in investigating a non-linear quantity,
namely the entropy,
\begin{equation}\label{entropy}
S = -{\rm Tr}\{\rho_{th}\ln\rho_{th}\} \quad ,
\end{equation}
and gauge how sensitive it is to experimental errors in $\left\langle {U_M } \right\rangle$. 

\begin{figure}[t]
\centering
\includegraphics[width= 0.9 \columnwidth]{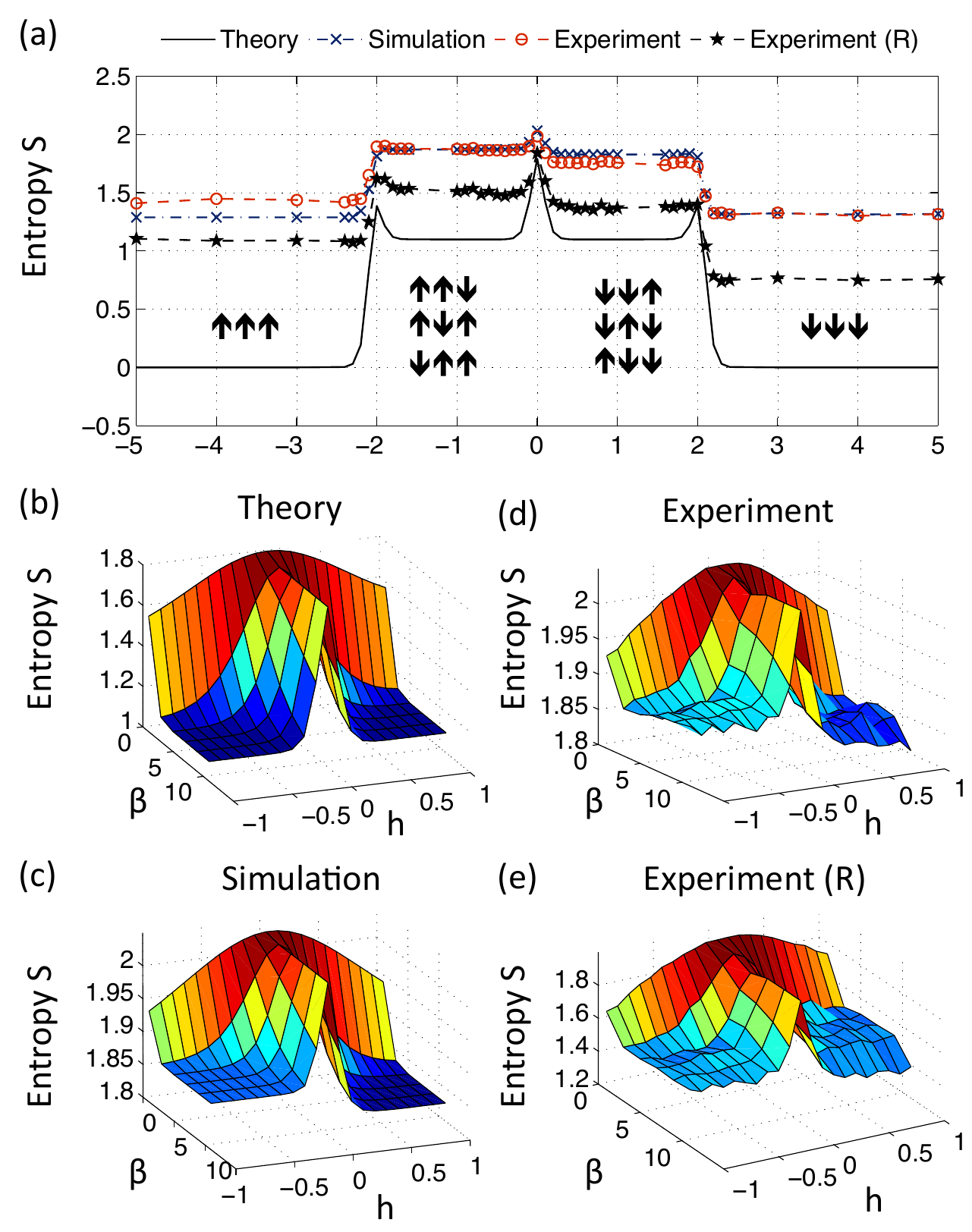} 
\caption{(a) Entropy $S$ as a function of magnetic field $h$ at
low temperature, $\beta=11$. The experimental data (``$\circ$") is
plotted together with numerical simulation results (``$\times$")
that include effects of carbon $T_2$ and proton $T_1$. The
theoretical result is shown as the solid curve. The sharp changes
of $S$ around $h=\pm2$ and $0$ indicate the phase transitions. The
points labeled ``Experiment (R)" are obtained from the
experimental data by using a simple decoherence model to partially
remove the effects of decoherence with no free parameters
(see SI). (b-e) Surface plots of entropy as a
function of $h$ and $\beta$ from theory (b), experiment (c) and
simulation (d). Modified experimental results that partially
remove decoherence effects are shown in (e).}
\label{Lowentropyfig}
\end{figure}
Fig.~\ref{Lowentropyfig}a shows the experimental results for the entropy~$S$ as a function of the simulated magnetic field $h$ in the low
temperature regime ($\beta = 11$). One finds that
the sharp changes of $S$ around $h=\pm2$ and $0$ correspond to the crossovers, which have been observed in measurement of the total magnetization. Compared with the region of $|h|>2$, the large values
of $S$ in the region $|h|<2$, especially around $h=0$, indicates the preference of anti-ferromagnetism which causes the
frustration for the magnet. For the outer region $|h|>2$, where the external field should be strong enough to polarize the magnet, the theoretical predictions of $S$ should be zero. However, the experiment results, including the rescaled results, show non-zero values. This is due to the anisotropy of the measurement results taken from the three nuclear spins. The surface plots in Fig.\ref{Lowentropyfig}b-d
show the entropy as a function of $h$ and $\beta$ from theory (b),
experiment (c), simulation (d), and the rescaled experimental results
that partially remove decoherence effects (e). We see that the role of the temperature is to ``wash out" the competition between the antiferromagnetic coupling and the external field. This is indicated by a transition near some value of the temperature ($\beta=3$), beyond which the variation of the simulated external field $h$ no longer causes sharp crossovers.

To summarize, the imperfection of the experimental results is mainly due to decoherence effects; the duration of the experiment
ranges from 0.35-0.76 s, (see supplementary material for details),
which is comparable with the $T_2$ times of the nuclear spins. The simulated data, which take into account the nuclear $T_2$ decay, closely match the experimental results.
To a much lesser extent, inhomogeneities of the applied magnetic fields and imperfect pulses also contribute to the error. Partial recovery
of the density matrix is possible for isotropic errors (see supplementary material), this allows us
to rescale the magnetization to improve the agreement with the
theoretical values (see Fig.~\ref{figmag}). The anisotropic part
of the error, however, cannot be removed, and significantly affects the experimental entropy results (see Fig.~\ref{Lowentropyfig}).

The phase kick-back method was chosen for readout to exploit the
fact that all J-couplings are well resolved for spin C$_1$. For
spins C$_2$ and C$_3$, there is non-negligible overlap between
certain spectral peaks, which makes direct state tomography
unreliable for certain states. Nonetheless, for certain simple
states like $|1111\rangle$ at at $\beta = 11$ and $h=5$, state
tomography through direct readout of all three spins can be
performed reliably and compared to the phase kick-back results.
The state fidelity measured in this fashion was close to the phase
kick-back result, only differing by $0.4\%$ (see Fig.
\ref{tomop5fig} in supplementary information).


In conclusion, we employed a quantum information processor using nuclear spins to perform a digital quantum simulation
of a geometrically frustrated magnet subject to a simulated magnetic field. We explored the phase diagram
of this system for a range of temperatures and magnetic fields, and studied the
competition between the antiferromagnetic couplings and the external field;
the crossover points where the magnetic field quenches the frustration are
correctly captured, and the overall experimental results are in good agreement
with numerical simulations and theory.

\section{Author Contributions} J.-F.Z. and J. B. designed the NMR experiments and simulations, which were carried out by J.-F.Z.; M.-H.Y. and A.A.-G. made the theoretical proposal and contributed to the analysis of results. R.L. and J. B. supervised the experiment. All authors contributed to the writing of the paper and discussed the experimental procedures and results.

\section{Acknowledgments}
We thank J. D. Whitfield for insightful discussions, and are
grateful to the following funding sources: Croucher Foundation (M.H.Y); DARPA under the Young Faculty Award
N66001-09-1-2101-DOD35CAP, the Camille and Henry Dreyfus
Foundation, and the Sloan Foundation; Army Research Office under
Contract No. W911NF- 07-1-0304 (A.A.G.); CIFAR, SHARCNET and QuantumWorks (R.L.), and NSERC (J.-F. Z., R. L. and J. B.)

\bibliographystyle{apsrev}

\clearpage
\begin{widetext}
\center
{\bf Supplementary information: Digital Quantum Simulation of the Statistical Mechanics of a Frustrated Magnet}
\medskip
\bigskip
\end{widetext}

\appendix

\section{Quantum circuit construction for creating the CETS}
For clarity, we re-write the Hamiltonian of the frustrated magnet as:
\begin{equation}
H = J\left( {s_1 s_2  + s_2 s_3  + s_1 s_3 } \right) + h\left( {s_1  + s_2  + s_3 } \right) \quad,
\end{equation}
where $s=\{0,1\}$, and $Z\left| s \right\rangle  = \left( { - 1} \right)^s \left| s \right\rangle$. To construct the quantum circuit diagram for creating the CETS of the frustrated magnet, we first consider the general property for the following controlled operation:
\begin{equation}\label{op_cRot}
\left| {s_1 s_2 } \right\rangle  \otimes \left| 0 \right\rangle  \to \left| {s_1 s_2 } \right\rangle  \otimes \left( {\cos \theta \left| 0 \right\rangle  + \sin \theta \left| 1 \right\rangle } \right) \quad,
\end{equation}
where
\begin{equation}\label{op_cos}
\cos \theta  = \sqrt {e^{ - \beta J\left( {s_1  + s_2 } \right) - \beta h} /K}
\end{equation}
and
\begin{equation}\label{op_sin}
\sin \theta  = \sqrt {e^{ + \beta J\left( {s_1  + s_2 } \right) + \beta h} /K} \quad,
\end{equation}
and
\begin{equation}
K \equiv e^{ - \beta J\left( {s_1  + s_2 } \right) - \beta h}  + e^{ + \beta J\left( {s_1  + s_2 } \right) + \beta h} \quad .
\end{equation}
Note that the numerators in Eq.~(\ref{op_cos}) and Eq.~(\ref{op_sin}) are chosen such that they give the correct weight of the Boltzmann factors. Now, one can show that
\begin{equation}\label{renorm_K}
K  = A e^{\beta b s_1 s_2 } e^{\beta c \left( {s_1  + s_2 } \right)} \, ,
\end{equation}
where
\begin{eqnarray}
 A  &=& 2\left( {\sqrt {\cosh \left( {2\beta J {+} \beta h} \right)\cosh \left( {2\beta J {-} \beta h} \right)} \cosh \left( {\beta h} \right)} \right)^{1/2} , \nonumber \\
 b  &=& \frac{1}{{4\beta }}\ln \left[ {\frac{{\cosh \left( {2\beta J + \beta h} \right)\cosh \left( {2\beta J - \beta h} \right)}}{{\cosh ^2 \left( {\beta h} \right)}}} \right] \, ,\nonumber \\
 c &=& \frac{1}{{4\beta }}\ln \left( {\frac{{\cosh \left( {2\beta J + \beta h} \right)}}{{\cosh \left( {2\beta J - \beta h} \right)}}} \right) \quad .
 \end{eqnarray}
Combining these results, we can interpret the operation in Eq.~(\ref{op_cRot}) as the one that gives the correct Boltzmann factors to the terms involving the third qubit, but it will renormalize the first two qubits, according to Eq.~(\ref{renorm_K}).

This suggests that in order to prepare the three-qubit CETS, we will need to prepare the CETS for the first two qubits with respect to a Hamiltonian which takes into account the renormalization effect caused by the third qubit:
\begin{equation}
H =  \tilde Js_1 s_2  + \tilde h\left( {s_1  + s_2 } \right) \quad,
\end{equation}
where
\begin{equation}
\tilde J \equiv J - b \quad,
\end{equation}
and
\begin{equation}
\tilde h \equiv h - c \quad.
\end{equation}
This can be achieved by the gates $U_x$, $U_y$ and $U_z$ shown in the quantum circuit diagram Eq.~(\ref{figcir})c. Their explicit forms are as follows: $U_x  \equiv R\left( {\theta _x } \right)$, $U_y  \equiv R\left( {\theta _y } \right)$, and $
U_z  \equiv R\left( {\theta _z } \right)R\left( {\theta _y } \right)^{ - 1}$, where
\begin{equation}
R\left( \theta  \right) = \left( {\begin{array}{*{20}c}
   {\cos \theta } & { - \sin \theta }  \\
   {\sin \theta } & {\cos \theta }  \\
\end{array}} \right) \quad ,
\end{equation}
and
\begin{equation}
\cos \theta _x  \equiv \sqrt {\frac{{e^{ - \beta \left( {\tilde h - g} \right)} }}{{2\cosh \left[ {\beta \left( {\tilde h - g} \right)} \right]}}} \quad,
\end{equation}
and
\begin{equation}
\cos \theta _y  \equiv \sqrt {\frac{{e^{ - \beta \left( {\tilde J + \tilde h} \right)} }}{{2\cosh \left[ {\beta \left( {\tilde J + \tilde h} \right)} \right]}}} \quad,
\end{equation}
and
\begin{equation}
\cos \theta _z  \equiv \sqrt {\frac{{e^{ - \beta \left( { - \tilde J + \tilde h} \right)} }}{{2\cosh \left[ {\beta \left( { - \tilde J + \tilde h} \right)} \right]}}} \quad.
\end{equation}
Here
\begin{equation}
g = \frac{1}{{2\beta }}\ln \frac{{\cosh \left( {\beta \tilde J + \beta \tilde h} \right)}}{{\cosh \left( {\beta \tilde J - \beta \tilde h} \right)}} \quad .
\end{equation}

The second part of the quantum circuit diagram implements the transformation Eq.~(\ref{op_cRot}), and involves the gates $U_0$, $U_1$ and $U_2$. Their explicit forms are given by: $U_0  \equiv R\left( {\theta _0 } \right)$, $U_1  \equiv R\left( {\theta _1 } \right)R\left( {\theta _0 } \right)^{ - 1}$, and $U_2  \equiv R\left( {\theta _2 } \right)T^{ - 1} \left( {\theta _1 ,\theta _0 } \right)$, where $T\left( {\theta _1 ,\theta _0 } \right) \equiv R\left( {\theta _1 } \right)R\left( {\theta _0 } \right)^{ - 1} R\left( {\theta _1 } \right)$,
\begin{equation}
\cos \theta _0  \equiv \sqrt {\frac{{e^{ - \left( {\beta 2J + \beta h} \right)} }}{{2\cosh \left( {\beta 2J + \beta h} \right)}}} \quad,
\end{equation}
and
\begin{equation}
\cos \theta _1  \equiv \sqrt {\frac{{e^{ - \beta h} }}{{2\cosh \left( {\beta h} \right)}}} \quad,
\end{equation}
and
\begin{equation}
\cos \theta _2  \equiv \sqrt {\frac{{e^{\beta 2J - \beta h} }}{{2\cosh \left( {\beta 2J - \beta h} \right)}}} \quad.
\end{equation}

\section{Partial recovery from isotropic noise} 
We applied an empirical transformation of the measurement results to partially remove decoherence errors. There is no rigorous proof that this transformation will work for all circumstances, however, there are reasonable motivations for it. In this experiment, all of the observables $\left\langle {U_M } \right\rangle$ are obtained by measuring the coherence of the probe qubit (see Eq. (\ref{sreduced})). For the moment, if we only take into account the $T_2$ decay of the probe qubit, then all of the measured values should be smaller than the actual value by a value roughly equal to $\eta \equiv e^{ - \tau /T_2 }$. Here $\tau$ is the evolution time, and we assume it is the same for all of the measurements.
This assumption is equivalent to saying that the prepared CETS $|\Psi_{\beta}\rangle$ is subject to the depolarizing channel:
\begin{equation}\label{depo_chan}
\rho _{\varepsilon}  \equiv \varepsilon \left( {\rho _{c} } \right) = \left( {1 - \eta } \right)\frac{I}{D} + \eta \rho _{c} \quad ,
\end{equation}
where $\rho _c  \equiv \left| {\Psi _\beta  } \right\rangle \left\langle {\Psi _\beta  } \right|$, and $D=8$ is the dimension of the CETS. 
In the ideal case where $\eta$ is known, the CETS can be perfectly recovered from $\rho _{\varepsilon}$ by the following transformation:
\begin{equation}
\rho _c  = \left( {\rho _\varepsilon   - \frac{I}{D}} \right) \times \frac{1}{\eta } + \frac{I}{D} \quad .
\end{equation}
Or equivalently, we may simply multiply all of the observables by the factor $\eta$.

In reality, a depolarizing channel as a noise model is a pure assumption, and it is almost impossible to determine the exact value of $\eta$, e.g. in the case where the decay rate for each observable may be different from the others. Nonetheless, partial recovery from the noise is possible when the decay rates are roughly the same, and we may estimate $\eta$ in an average sense. 
To see this, consider the trace-preserving transformation:
\begin{equation}
\left( {\rho _\varepsilon   - \frac{I}{D}} \right) \times \frac{1}{\lambda } + \frac{I}{D} = \frac{\eta }{\lambda }\rho _c  + \left( {1 - \frac{\eta }{\lambda }} \right)\frac{I}{D} = \rho'
\end{equation}
where $\lambda$ is assumed to be close to $\eta$. It is easy to check that this transformation is not positive. However, as long as $\lambda$ is close to $\eta$, the rescaled state $\rho'$ is approximately equal to $\rho_c$, i.e., the original CETS. 

In fact, for the depolarizing channel applied to a pure state, i.e., $Tr(\rho_c^2)=1$, a systematic estimation for $\eta$ is possible. To see this, consider taking the trace of the square of $\rho_{\varepsilon}$ in Eq.~(\ref{depo_chan}), we find that
\begin{equation}\label{exact_eta}
\eta  = \sqrt {\frac{{Tr\left( {\rho _\varepsilon ^2 } \right) - 1/D}}{{1 - 1/D}}} \quad.
\end{equation}
When $D$ is much larger than $1$, we have the approximation
\begin{equation}
\eta  \approx \sqrt {Tr\left( {\rho _\varepsilon ^2 } \right)} \quad.
\end{equation}
If we take this correction to the density matrix to calculate the fidelity with the $\left| {\Psi _\beta  } \right\rangle$ CETS, the result, in the limit $D \gg 1$, is the same as that of the projection between them:
\begin{equation}
P \equiv \frac{{\left\langle {\Psi _\beta  } \right|\rho _\varepsilon  \left| {\Psi _\beta  } \right\rangle }}{{\sqrt {Tr\left( {\rho _\varepsilon ^2 } \right)} }} \quad.
\end{equation}

Although, in theory, Eq. (\ref{exact_eta}) is an exact expression for $\eta$, in practice, it can easily cause the final density matrix to have negative eigenvalues. Empirically, the approximation $\eta  \approx \sqrt {Tr\left( {\rho _\varepsilon ^2 } \right)}$ works much better. In this experiment, we chose $\eta = 0.6316$, which is equal to the square root of the purity of the measured CETS at $\beta =11$ and $h=5$. With a single fitting parameter, the total magnetizations for all of the states are rescaled to values much closer to the theoretical values, as shown in Fig.~\ref{figmag}a-d.

The term ``isotropic error" in the main text refers to the part of the error generated by the uniform part of the decoherence. To make the statement more quantitative, as an example, consider two observables which suffer from two different dephasing rates $\gamma_1$ and $\gamma_2$. The isotropic error can be quantified by defining the mean value $\gamma _m  \equiv \left( {\gamma _1  + \gamma _2 } \right)/2$. By rescaling the factor $e^{\gamma _m t}$ to both observables, the effective decay rate becomes the anisotropic error rates $\gamma _1  - \gamma _m$ and $\gamma _2  - \gamma _m$, which is zero when $\gamma_1 = \gamma_2$.

Lastly, we emphasize again that this decoherence model is a pure assumption, and to motivate the use of it, we considered only the dephasing of the probe qubit. In reality, there are also decoherence channels for the other qubits. On the other hand, the depolarizing noise model is a useful approximation when we consider that the applied pulse sequences may randomize the noise to some extent. In the limit that we apply a large number of random operations uniformly drawn from the Clifford gates, we would in fact get exactly a uniform depolarizing channel (See, e.g., C A Ryan et al 2009 New J. Phys. 11 013034).

\begin{figure*}[t]
\includegraphics[width=0.7 \textwidth]{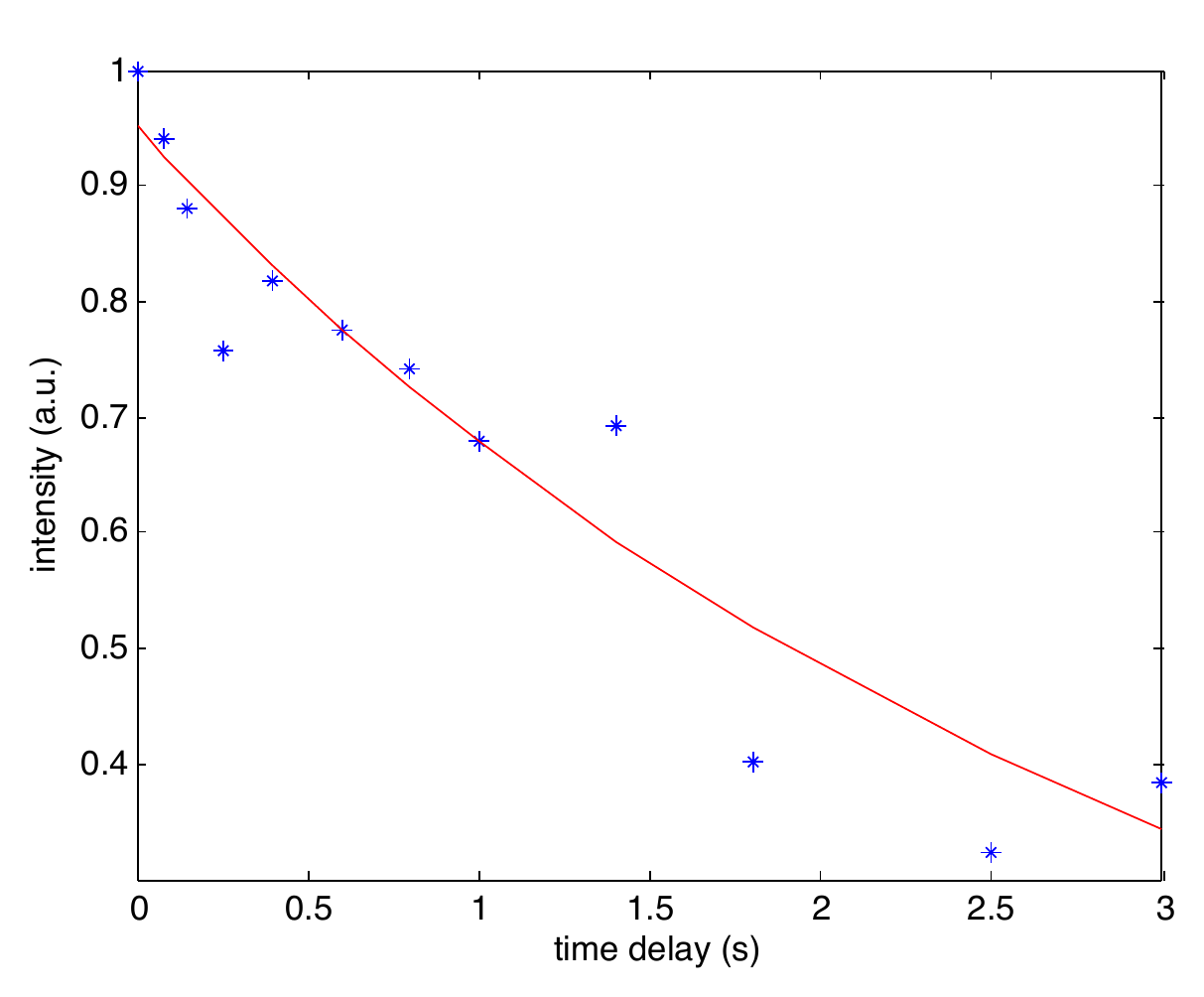}
\caption{The decay of the H$_2$ signal obtained by a $\pi/2$
readout pulse to the state
$\mathbf{0}\mathbf{0}\sigma^{z}\one\one\one\one$. The
experimental  data are denoted by $*$, and the fitting
result is shown as the solid curve. }. \label{figT1}
\end{figure*}

\section{$T_1$ relaxation of the proton spins}
In implementating the pseudopure state, $\rho_{s}$ starts from the
state $\sigma^{z}\one\one\one\one\one\one$.

In the experiment, the proton spins M, H$_1$ and H$_2$ are prepared in the state
$\mathbf{0}\mathbf{0}\sigma^{z}$.  We measured the decay
of the state $\mathbf{0}\mathbf{0}\sigma^{z}\one\one\one\one$ in order to
estimate the effect of the protons' $T_1$ relaxation process on the measurement of the CETS. The state $\mathbf{0}\mathbf{0}\sigma^{z}\one\one\one\one$
is prepared from $\rho_{s} = \sigma^{z}\one\one\one\one\one\one$ through phase cycling. After a delay time, a $\pi/2$ readout pulse for H$_2$ is applied.  The intensity of the signal against the
delay time is shown in Fig. \ref{figT1}, with a fit yielding a relaxation time of 2.95 s.

\section{Supplementary data}
The supplementary figures are shown below.
They include a table summarizing the experimental durations for various measurement observables,
the spectra of the probe qubit, and selected measurement results not included in the main text.

\begin{table*}[h]  
\begin{tabular}{|c|c|c|c|}
\hline  Measurement  & Experiment duration (s) & Theoretical value  & Experimentally measured value\\
\hline  $\langle Z_1\rangle$   & 0.35 & -1/3  & -0.2497\\
\hline   $\langle Z_2\rangle$   &  0.46 & -1/3  & -0.1208 \\
\hline  $\langle Z_3\rangle$  & 0.57 & -1/3 & -0.2564  \\
 \hline $\langle Z_1Z_2\rangle$  & 0.62 & -1/3 & -0.1335 \\
\hline  $\langle Z_2Z_3\rangle$  & 0.58 &  -1/3 & -0.3377\\
\hline  $\langle Z_1Z_3\rangle$  &  0.76 & -1/3 & -0.1549\\
\hline  $\langle Z_1Z_2Z_3\rangle$   & 0.59 & 1 & 0.5110 \\
\hline  $\langle X_1\rangle$  & 0.37  &0 & 0.2007\\
\hline  $\langle X_2\rangle$  & 0.49 & 0& 0.0743\\
\hline   $\langle X_3\rangle$  &  0.63 & 0 & 0.0154   \\
\hline $\langle Y_1\rangle$  & 0.37 & 0 & -0.1261 \\
\hline  $\langle Y_2\rangle$  & 0.49 & 0 & -0.1375 \\
\hline  $\langle Y_3\rangle$  &  0.63 & 0 & -0.1769\\
\hline
\end{tabular}
 \caption {Experimental parameters for estimating errors in implementation. Experiment duration for measuring $U_M$
  includes the preparation of CETS, which takes $0.33$ s. The theoretical and experimental data correspond to
   $T = 1/11$ and $h = 1$. The differences in the errors in for $U_M$ give rise to the anisotropic part of the error in the reconstructed density matrix, which cannot be removed by the empirical transformation described above.}  \label{duration}
\end{table*}

\begin{figure*}[h]
\includegraphics[width= 0.8 \textwidth]{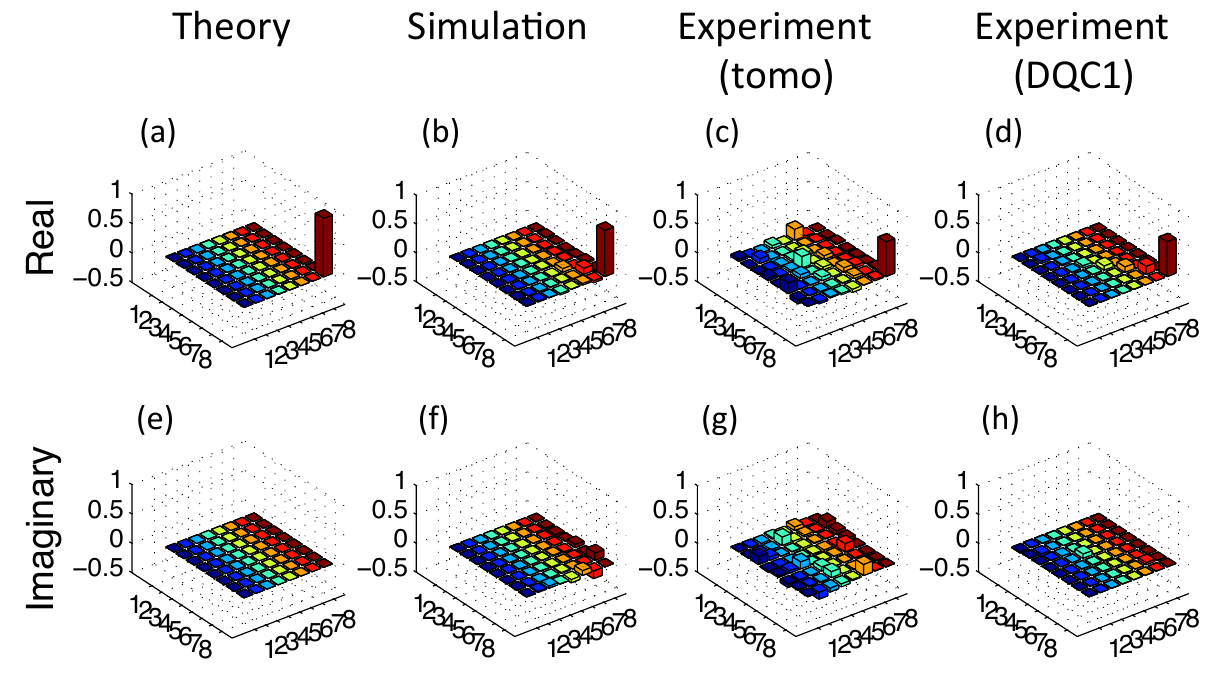} 
\caption{Density matrices for $T = 1/11$ and $h = 5$. From left to
right, the four columns show the results in theory, by simulation (including $T_2$),
in experiment by full state tomography \cite{PRA04}, and in experiment using the phase kick-back readout. The top and bottom rows show the real and imaginary
parts, respectively. Compared with the theoretical result, the state fidelities from the simulation, state tomography and phase kick-back method are $79.8\%$, $60.2\%$ and $59.8\%$,
respectively.} \label{tomop5fig}
\end{figure*}

\begin{figure*}[h]
\includegraphics[width=0.5 \textwidth]{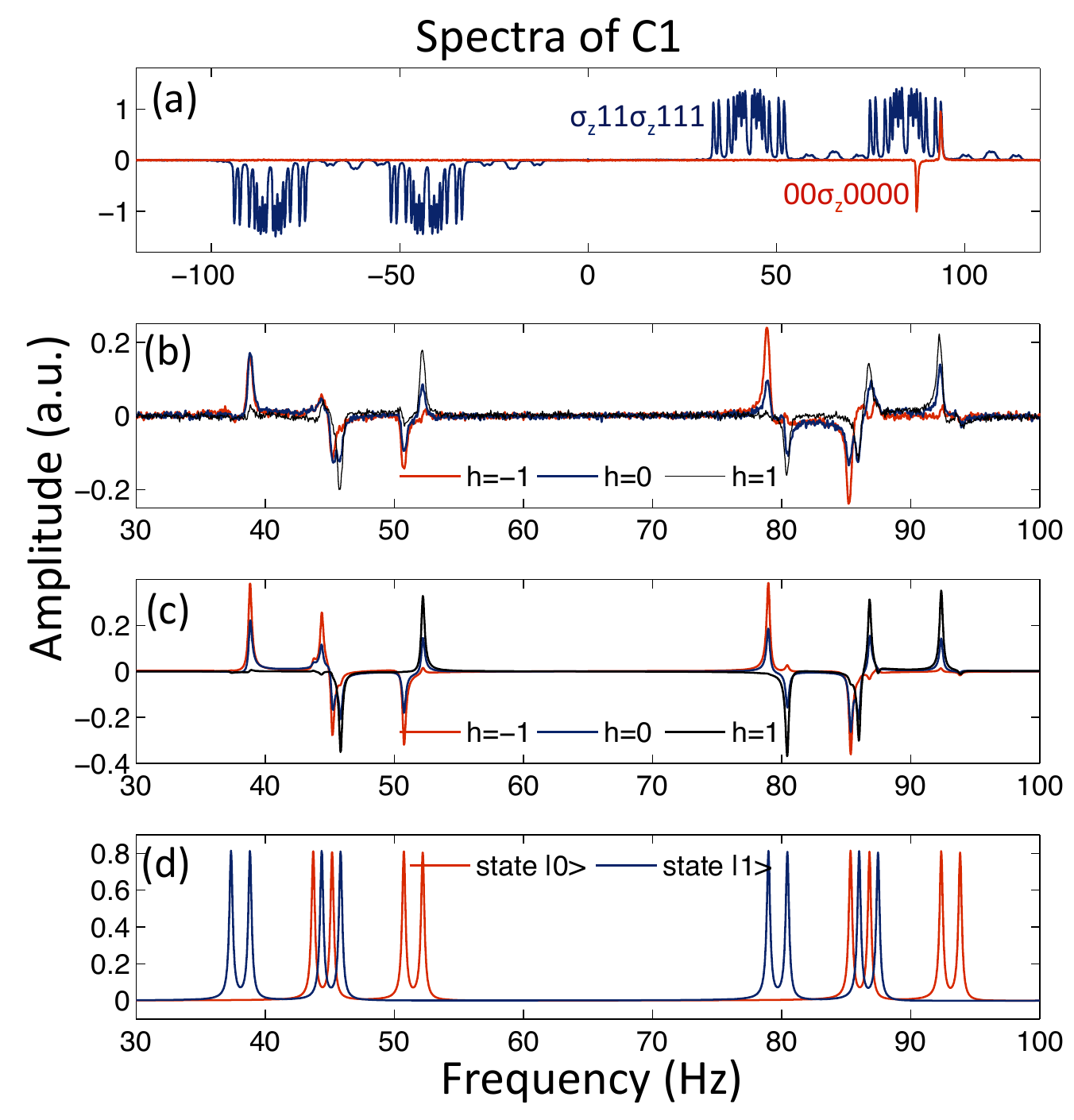}
\caption{ (a) Spectra of C$_1$ obtained by $\pi/2$ readout pulses
when the system is in the labelled pseudo-pure state
$\rho_{s}=\mathbf{0}\mathbf{0}\sigma_z
\mathbf{0}\mathbf{0}\mathbf{0}\mathbf{0}$ (red) and reference
state $\sigma_z\one\one\sigma_z\one\one\one$ (blue), respectively.
The vertical axes have the same scaling (arbitrary units). (b,c)
NMR spectra of C$_1$ for measuring $\langle Z_1Z_2Z_3\rangle$ in
experiment (b) and by simulation that includes $T_2$ effects (c),
with $\beta=5$, $h=-1$ (red), $0$ (blue), and $1$ (black). In each
spectrum, only the peaks marked by $M$ in state $|0\rangle$ are
shown, because the peaks marked by $M$ in state $|1\rangle$ are
too weak to attribute observable signals). (d) For reference,
simulated spectra of C$_1$ marked by spin H$_2$ in state
$|0\rangle$ (dashed) and $|1\rangle$ (solid) are shown (with
C$_2$-C$_4$  in the state $\one^{\otimes 3}$).}. \label{figpps}
\end{figure*}

\begin{figure*}[h]
\includegraphics[width= 0.8 \textwidth]{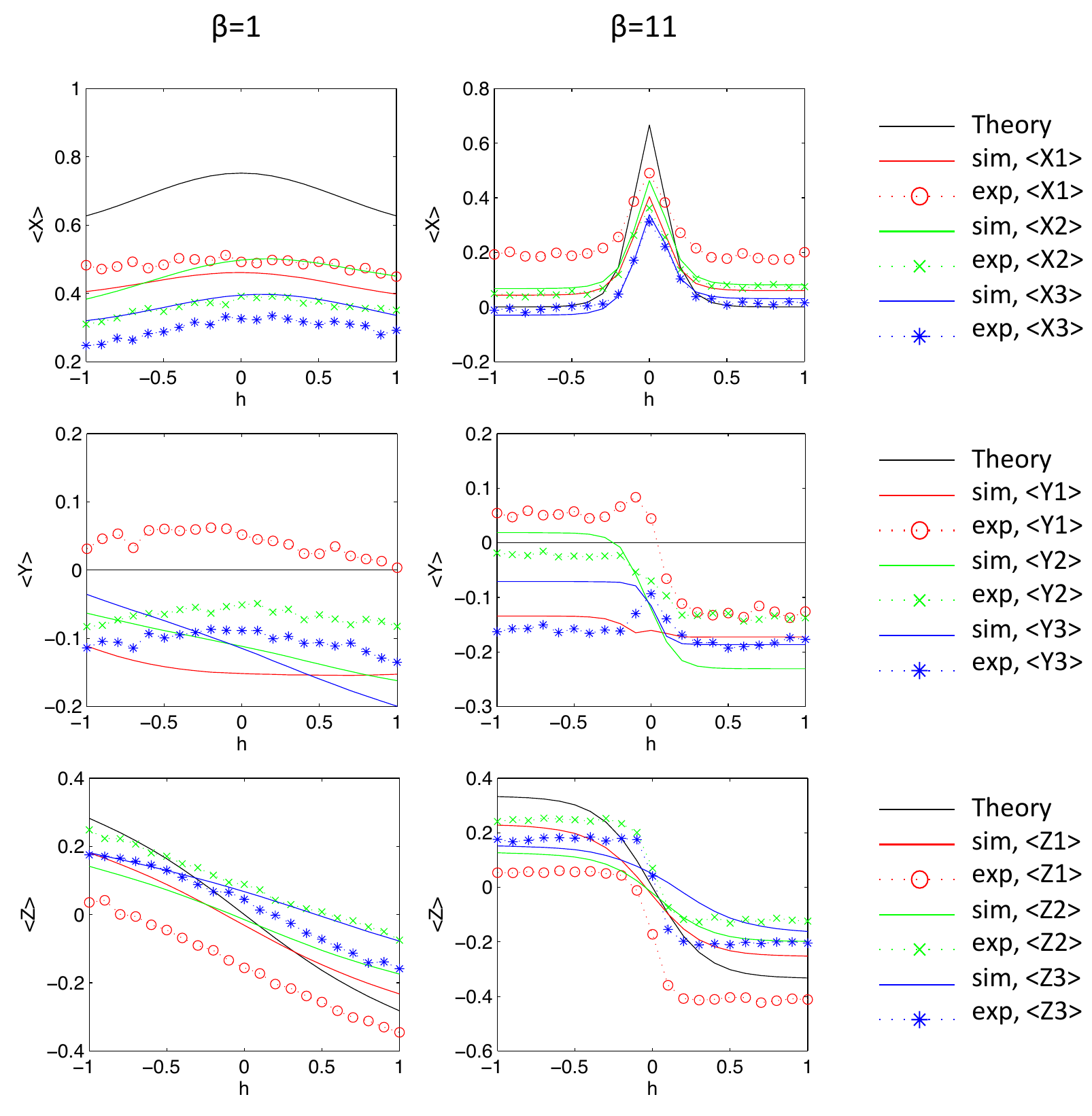}
\caption{Measurement results of $\langle X \rangle$, $\langle Y \rangle$ and $\langle Z \rangle$ for the high-temperature limit $\beta=1$ (left) and the low-temperature limit $\beta=11$ (right).}
\label{Xfig}
\end{figure*}

\end{document}